\documentclass[a4paper,11pt]{article}
\usepackage[dvips]{graphicx}
\usepackage{amssymb}
\usepackage{amsmath}

\usepackage{cite}

\input arrow

\begin{document}

\title{A Relation Between $Z_3$-Graded Symmetry and Shape Invariant Supersymmetric Systems}
\author{
V.K. Oikonomou$^{1,}$$^{2}$\,\thanks{voiko@physics.auth.gr}\\
$^1$Department of Mechanical Engineering, Technological Education Institute of Serres\\
62124 Serres, Greece \\
$^2$Department of Theoretical Physics, Aristotle University of Thessaloniki,\\
54124 Thessaloniki, Greece
} \maketitle

\begin{abstract}
We study an interesting property of shape invariant supersymmetric quantum mechanical systems. Particularly, we demonstrate that each shape invariant supersymmetric system can constitute a $Z_3$-graded topological symmetric algebra. The latter is known to provide topological invariants which are generalizations of the Witten index. In addition, we relate the $Z_3$-graded algebra to the generators of the $SO(2,1)$ Lie algebra underlying each shape invariant system. We generalize the results to the case of sequential shape invariant systems, in which case we find a sequence of $Z_3$-graded algebras. Finally, we briefly discuss two systems that are related to shape invariance, but have different algebraic origin, namely supersymmetric systems with central charge equipped with an additional symmetry and $Z_3$-graded algebraic systems. In view of the fact that the shape invariance condition is somewhat an additional algebraic condition, with no origin to some concrete algebraic structure, our results might be useful towards this line of research.
\end{abstract}

\section*{Introduction}

Supersymmetry was initially introduced in quantum field theory as a graded Lie extension of the four dimensional Poincare algebra \cite{susy1}. This super-Poincare algebra relates fermionic and bosonic representations in a direct way and consequently, every boson has its supersymmetric fermionic counter partner. However, as the current experiments indicate, supersymmetry has to be broken in our four dimensional world. There are various ways to break supersymmetry, even in the context of grand unified theories \cite{susy1,odi1,odi2} and the breaking has to be somehow controlled in order to imitate the phenomenological constraints. Supersymmetry offers many elegant qualitative features to a field theory \cite{susy1,odi1,odi2}, such  as the absence of quadratic divergences and also is desirable to most of grand unified field theories and sting theories. In addition, it offers many good features to cosmological and supergravity theories \cite{odi3}. Supersymmetric quantum mechanics (SUSY QM hereafter), was introduced by Witten to address the issue of supersymmetry breaking in quantum field theory \cite{witten1}. Nowadays, SUSY QM is a powerful tool for studying dimensionally reduced quantum field theories and integrability in quantum mechanical systems. For detailed reviews and textbooks on SUSY QM, see for example \cite{reviewsusyqm} and references therein. In this line of research, shape invariance \cite{reviewsusyqm,shapeinvariancebasic,shapeinvarianceapplications}, is a basic tool for finding in a simple and concrete way solutions to SUSY QM related quantum mechanical systems, and therefore is crucial for the integrability of certain systems. SUSY QM is a field of research which is by itself interesting and is not only a one dimensional theoretical tool for studying in a simple way higher dimensional quantum field theories. It was soon realized that SUSY QM provides insights to the factorization method \cite{reviewsusyqm}. The factorization method is used to categorize the analytically solvable potential problems and is closely related to shape invariance \cite{reviewsusyqm,shapeinvariancebasic,shapeinvarianceapplications}. The applications of SUSY QM are numerous covering a wide range of research areas, for example mathematical properties of SUSY QM Hilbert spaces and also applications to  studies of quantum mechanical systems. Studies of extended supersymmetries and harmonic superspaces or gravity, were done in \cite{extendedsusy,ivanov}, while scattering applications of SUSY QM can be found in \cite{susyqmscatter}. Applications in quantum mechanical systems and interesting features of supersymmetry breaking can be found in \cite{susyqminquantumsystems} and \cite{susybreaking} respectively. In addition, some geometrical and extended SUSY QM applications of SUSY QM methods can be found in \cite{diffgeomsusyduyalities} and \cite{plu1,plu3,plu4}. Regardless the fact that SUSY QM and global four dimensional spacetime supersymmetry are related for some theories \cite{ivanov}, in principle these two concepts are completely different at least conceptually.

In this article, the focus is on a particular property of every shape invariant SUSY QM system, with unbroken supersymmetry. Specifically we shall demonstrate that each shape invariant SUSY QM system can constitute a $Z_3$-graded symmetric topological quantum mechanical system \cite{ali1,ali2}. Topological symmetries in the spirit of references \cite{ali1,ali2}, are symmetries that generalize the concept of the Witten index and therefore provide useful insights to further develop the algebraic structure of SUSY QM systems. We shall establish the result that shape invariant systems can constitute such $Z_3$-graded systems and therefore, since shape invariance is actually imposed as an ad-hoc relation without having a deeper structural-algebraic reason, our results might shed some light towards the problem of finding a deeper explanation of the occurrence of shape invariance. A generalization to sequential multi-shape invariant SUSY QM systems \cite{bazeia} follows. In addition, since every shape invariant system is related to an inherent $SO(2,1)$ symmetry \cite{bala1}, we shall express the $Z_3$-graded symmetric system in terms of the generators of the $SO(2,1)$ Lie algebra. Finally, we briefly discuss the similarities of central charge extended SUSY QM systems with an additional extra symmetry \cite{spector}, to the $Z_3$-graded symmetric SUSY QM systems, using the shape invariance as a common feature between the two concepts. We believe our results offer some new insights to further understand the algebraic origin of shape invariance.

This paper is organized as follows: In section 1, we discuss the general framework of $Z_3$-graded symmetric systems, providing a brief introduction to the basic features that we will use. Moreover, we study how each shape invariant SUSY QM system can be used to construct a $Z_3$ symmetric system. Finally, we generalize our results to the case of multi-shape invariant systems and we find how multi-$Z_3$ symmetric systems are constructed. In section 2, we express the $Z_3$-graded symmetric structure in terms of the generators of the inherent $SO(2,1)$ Lie algebra. Additionally, we briefly discuss the possible algebraic connection of the $Z_3$ algebraic symmetry to SUSY QM systems with an additional symmetry. The conclusions follow in the end of the article.

\section{An Inherent $Z_3$-graded Symmetric Structure Underlying Shape Invariant Supersymmetric Systems}

In this section we shall demonstrate that each shape invariant supersymmetric quantum system can constitute a $Z_3$-graded system. In order to make the article self-contained, we shall present the necessary information on the $Z_3$ symmetry, following \cite{ali1,ali2}.

\subsection{$Z_3$-graded Topological Symmetry}

A $Z_3$-graded topological symmetry of type $(1,1,1)$ is defined using a grading operator $\tau$, which is related to the third root of unity, and satisfies the following relations:
\begin{equation}\label{go}
\tau^3=1,{\,}{\,}{\,}\tau^{\dag}=\tau^{-1},{\,}{\,}{\,}[H,\tau ]=0,{\,}{\,}{\,}[\tau,\mathcal{Q}]_q=0
\end{equation}
with $\mathcal{Q}$, the generator of the topological symmetry $q=e^{2\pi i/3}$ and the commutator $[,]_q$ stands for the $q$-commutator,
\begin{equation}\label{qcom}
[O_1,O_2]_q=O_1O_2-qO_2O_1
\end{equation}
The operator algebra for $Z_3$-graded topological symmetry of type $(1,1,1)$ has the form:
\begin{align}\label{topolsymmetrop}
&\mathcal{Q}^3=\mathcal{K} \\ \notag & 
\mathcal{Q}_1^3+\mathcal{M}\mathcal{Q}_1=2^{-3/2}(\mathcal{K}+\mathcal{K}^{\dag}),\\ \notag &
\mathcal{Q}_2^3+\mathcal{M}=-2^{-3/2}i(-\mathcal{K}+\mathcal{K}^{\dag})\\ \notag &
[\mathcal{M},\mathcal{Q}]=0,{\,}{\,}{\,}[K,\mathcal{Q}]=0
\end{align}
with $\mathcal{K}$ and $\mathcal{M}$ operators, the commutator of which with all the other operators is zero. We shall take $\mathcal{K}=H$, with $H$, the Hamiltonian of the quantum system. In addition, the operator $\mathcal{M}$ is self adjoint and its specific form imposes some restrictions on the system as we shall see. The operators $\mathcal{Q}_1$ and $\mathcal{Q}_2$ are defined in terms of the operator $\mathcal{Q}$, as follows,
\begin{equation}\label{q}
\mathcal{Q}_1=\frac{\mathcal{Q}+\mathcal{Q}^{\dag}}{\sqrt{2}},{\,}{\,}{\,}\mathcal{Q}_2=\frac{\mathcal{Q}-\mathcal{Q}^{\dag}}{\sqrt{2}i}
\end{equation}
A quantum system with a Hamiltonian $H$ has a $Z_3$-graded topological symmetry of type $(1,1,1)$ if there exist operators $\tau$ and $\mathcal{Q}$ satisfying (\ref{go}), the spectrum of the Hamiltonian is non-negative and finally $\mathcal{Q}^3=H$. We shall use a three dimensional representation of the algebra (\ref{topolsymmetrop}). Consider three Hilbert spaces $\mathcal{H}_1,\mathcal{H}_2,\mathcal{H}_3$ and their direct sum $\mathcal{H}$. Let the vectors $|\psi _i\rangle$, $i=1,2,3$, belong to $\mathcal{H}_i$. In the three dimensional representation, the vector $|\psi \rangle$, being a member of the total Hilbert space $\mathcal{H}$, can be represented as:
\begin{equation}\label{psiol}
|\psi \rangle =\left ( \begin{array}{c}
  |\psi_1 \rangle \\
  |\psi_2 \rangle \\
|\psi_3 \rangle \\
\end{array} \right)
\end{equation}
In the same representation, the grading operator $\tau$, is defined by,
\begin{equation}\label{grad}
\tau= \left ( \begin{array}{ccc}
  q & 0 & 0  \\
  0 & q^2 & 0  \\
0 & 0 & 1 \\
\end{array} \right)
\end{equation}
while the operators self-adjoint operators $\mathcal{Q}$ and $\mathcal{M}$, are equal to:
\begin{equation}\label{grad1}
\mathcal{Q}= \left ( \begin{array}{ccc}
  0 & 0 & \mathcal{D}_3  \\
  \mathcal{D}_1 & 0 & 0  \\
0 & \mathcal{D}_2 & 0 \\
\end{array} \right),{\,}{\,}{\,}\mathcal{M}=\left ( \begin{array}{ccc}
  \mathcal{M}_1 & 0 & 0  \\
  0 & \mathcal{M}_2 & 0  \\
0 & 0 & \mathcal{M}_3 \\
\end{array} \right)
\end{equation}
The operators $\mathcal{D}_i$ are defined to be maps between the various Hilbert spaces. More specifically,
\begin{equation}\label{o}
\mathcal{D}_1:\mathcal{H}_1\rightarrow \mathcal{H}_2,{\,}{\,}{\,}\mathcal{D}_2:\mathcal{H}_2\rightarrow \mathcal{H}_3,{\,}{\,}{\,}\mathcal{D}_3:\mathcal{H}_3\rightarrow \mathcal{H}_1
\end{equation}
The operators $\mathcal{M}_i$ are automorphisms of the same Hilbert space, that is $\mathcal{M}_i:\mathcal{H}_i\rightarrow \mathcal{H}_i$. The Hamiltonian $H$, is represented by the following operator:
\begin{equation}\label{fghg}
H=\left ( \begin{array}{ccc}
  H_1 & 0 & 0  \\
  0 & H_2 & 0  \\
0 & 0 & H_3 \\
\end{array} \right)
\end{equation}
with the operators $H_i$ being the Hamiltonians corresponding to the Hilbert spaces $\mathcal{H}_i$. The conditions $[H,\mathcal{Q}]=0$, $[\mathcal{Q},\mathcal{M}]=0$ and the fact that each $H_i$ is self adjoint, impose some restrictions on the operators $\mathcal{D}_i$. Indeed, using the representations (\ref{grad1}), we get the following compatibility conditions:
\begin{align}\label{cond1}
&\mathcal{D}^{\dag}_1\mathcal{D}^{\dag}_2\mathcal{D}^{\dag}_3=\mathcal{D}_3\mathcal{D}_2\mathcal{D}_1 \\ \notag &
\mathcal{D}^{\dag}_2\mathcal{D}^{\dag}_3\mathcal{D}^{\dag}_1=\mathcal{D}_1\mathcal{D}_3\mathcal{D}_2 \\ \notag &
\mathcal{D}^{\dag}_3\mathcal{D}^{\dag}_1\mathcal{D}^{\dag}_2=\mathcal{D}_2\mathcal{D}_1\mathcal{D}_3
\end{align}
In addition, using the three dimensional representation for $\mathcal{Q}$ and relation (\ref{q}), we get the following additional set of compatibility conditions:
\begin{align}\label{cond2}
&\mathcal{D}^{\dag}_2\mathcal{D}_2\mathcal{D}_1\mathcal{D}_3=\mathcal{D}_1\mathcal{D}_3\mathcal{D}_2\mathcal{D}_2^{\dag} \\ \notag &
\mathcal{D}^{\dag}_3\mathcal{D}_3\mathcal{D}_2\mathcal{D}_1=\mathcal{D}_2\mathcal{D}_1\mathcal{D}_3\mathcal{D}_3^{\dag} \\ \notag &
\mathcal{D}^{\dag}_1\mathcal{D}_1\mathcal{D}_3\mathcal{D}_2=\mathcal{D}_3\mathcal{D}_2\mathcal{D}_1\mathcal{D}_1^{\dag}
\end{align}
Obtaining a solution that simultaneously satisfies relations (\ref{cond1}) and (\ref{cond2}), the quantities defined below are topological invariants:
\begin{align}\label{ti}
& \Delta_{12}=-\Delta_{21}=\mathrm{dim(ker}\mathcal{D}_1\mathcal{D}_3\mathcal{D}_2)-\mathrm{dim(ker}\mathcal{D}_3\mathcal{D}_2\mathcal{D}_1)\\ \notag & \Delta_{23}=-\Delta_{32}=\mathrm{dim(ker}\mathcal{D}_2\mathcal{D}_1\mathcal{D}_3)-\mathrm{dim(ker}\mathcal{D}_1\mathcal{D}_2\mathcal{D}_2)
\Delta_{13}=-\Delta_{31}=\Delta_{12}+\Delta_{23}
\end{align} 
In reference \cite{ali1} the author gives a non-trivial solution satisfying the constraints (\ref{cond1}) and (\ref{cond2}), which apply for arbitrary choices of the various Hilbert spaces $\mathcal{H}_i$. This solution corresponds to the choice $\mathcal{D}_3=\mathcal{D}_1^{\dag}\mathcal{D}_2^{\dag}$, which is automatically satisfied when the following relation is satisfied,
\begin{equation}\label{br}
[\mathcal{D}_1\mathcal{D}_1^{\dag},\mathcal{D}_2^{\dag}\mathcal{D}_2]=0
\end{equation}
Then the Hamiltonians $H_1,H_2,H_3$ of the quantum system is trivially given by:
\begin{equation}\label{ham}
H_1=\mathcal{D}_3^{\dag}\mathcal{D}_3,{\,}{\,}{\,}H_2=\mathcal{D}_1\mathcal{D}_1^{\dag}\mathcal{D}_2^{\dag}\mathcal{D}_2,{\,}{\,}{\,}H_3=\mathcal{D}_3\mathcal{D}_3^{\dag}
\end{equation}

\subsection{Description of the Inherent $Z_3$-graded Topological Structure for Each Shape Invariant Quantum Systems}

In this paper the focus is on revealing a very general system that can constitute a $Z_3$-graded topological symmetric quantum system, satisfying relations (\ref{cond1}), (\ref{cond2}) and (\ref{br}). As we shall demonstrate, all the shape invariant supersymmetric quantum mechanical systems have an inherent $Z_3$-graded topological structure. To start with, let us recall the definition of shape invariant supersymmetric systems and introduce some notation. For conventional Hermitian Hamiltonians $H(x,a)$, the SUSY QM problem is described by the operators $\mathcal{A}(x,a)$ and $\mathcal{A}^{\dag}(x,a)$. In the previous, $x$ denotes the space coordinate, while $a$, a general parameter of the problem. The SUSY QM algebra, due to the corresponding involution operator, provides the total Hilbert space $\mathcal{H}_{tot}$ corresponding to the Hamiltonian $H(x,a)$ with a $Z_2$ grading. This grading splits the total Hilbert space as follows,
\begin{equation}\label{gradddf}
\mathcal{H}_{tot}(x,a)=\mathcal{H}_{-}(x,a)\oplus \mathcal{H}_{+}(x,a)
\end{equation}
The Hamiltonian of the whole systems is split accordingly as follows:
\begin{equation}\label{splitham}
H(x,a)=H_{+}(x,a)+H_{-}(x,a)
\end{equation}
The sub-Hamiltonians are written in terms of the operators $\mathcal{A}(x,a)$ and $\mathcal{A}^{\dag}(x,a)$, 
\begin{equation}\label{splitham111}
H_{+}(x,a)=\mathcal{A}(x,a)\mathcal{A}^{\dag}(x,a),{\,}{\,}{\,}H_{-}(x,a)=\mathcal{A}^{\dag}(x,a)\mathcal{A}(x,a)
\end{equation}
Most importantly, let us note that the Hilbert space corresponding to the eigenstates of the Hamiltonian $H_{+}(x,a)$ is $\mathcal{H}_{+}(x,a)$, while the Hilbert space corresponding to the eigenstates of $H_{-}(x,a)$ is $\mathcal{H}_{-}(x,a)$. We made use of the parameter $a$ because we want to discriminate between shape invariant Hamiltonians. The shape invariant quantum systems correspond to different values of the parameter $a$, and the condition that ensures shape invariance for two quantum systems is:
\begin{equation}\label{sic}
\mathcal{A}(x,a_0)\mathcal{A}^{\dag}(x,a_0)-\mathcal{A}^{\dag}(x,a_1)\mathcal{A}(x,a_1)=f(a_0)
\end{equation}
with the parameters $a_0$ and $a_1$ characterizing the two different quantum systems. There are various types of shape invariance, like translational or scaling invariance, but our results do not depend on this. To make our arguments clear, let us show how the various operators of the two shape invariant systems act as maps between the corresponding Hilbert spaces. The operators $\mathcal{A}_i$ act as follows:
\begin{align}\label{mapact}
& \mathcal{A}(x,a_0): \mathcal{H}_{-}(x,a_0)\rightarrow \mathcal{H}_{+}(x,a_0) \\ \notag &
\mathcal{A}^{\dag}(x,a_0): \mathcal{H}_{+}(x,a_0)\rightarrow \mathcal{H}_{-}(x,a_0) \\ \notag &
\mathcal{A}(x,a_1): \mathcal{H}_{-}(x,a_1)\rightarrow \mathcal{H}_{+}(x,a_1) \\ \notag &
\mathcal{A}^{\dag}(x,a_1): \mathcal{H}_{+}(x,a_1)\rightarrow \mathcal{H}_{-}(x,a_1) \\ \notag &
\end{align}
Owing to the shape invariance conditions, the Hamiltonians and correspondingly the Hilbert spaces of the shape invariant systems are related. Thus, we may make the following identifications
\begin{equation}\label{ident1}
\mathcal{H}_{-}(x,a_0)\equiv \mathcal{H}_{+}(x,a_1),{\,}{\,}{\,}\mathcal{H}_{+}(x,a_0)\equiv \mathcal{H}_{-}(x,a_1)
\end{equation}
Keeping this identification in mind, we may construct the $Z_3$-graded quantum system using the operators $\mathcal{A}(x,a_0)$ and $\mathcal{A}(x,a_1)$. Recall the operators $\mathcal{D}_1,\mathcal{D}_2,\mathcal{D}_3$ we introduced in relation (\ref{grad1}). If we make the following identifications, the $Z_3$-graded structure naturally follows (by making only one assumption as we shall see):
\begin{align}\label{iddfhf}
&\mathcal{D}_1=\mathcal{A}(x,a_1)\\ \notag &
\mathcal{D}_2=\mathcal{A}(x,a_0) \\ \notag &
\mathcal{D}_3=\mathcal{D}_1^{\dag}\mathcal{D}_2^{\dag}
\end{align}
The $Z_3$ structure is guaranteed only if we make the assumption that the Hilbert space $\mathcal{H}_3$ is actually $\mathcal{H}_1$, or equivalently that the operator $\mathcal{D}_3$ of the $Z_3$ quantum system is actually an automorphism of the Hilbert space $\mathcal{H}_1$. To reveal the $Z_3$ structure of the shape invariant quantum system, notice the way the operators $\mathcal{D}_i$ as identified from relation (\ref{iddfhf}) act as maps between the Hilbert spaces, namely:
\begin{align}\label{atlanteansword}
&\mathcal{D}_1=\mathcal{A}(x,a_1): \mathcal{H}_{-}(x,a_1)\equiv \mathcal{H}_1\rightarrow \mathcal{H}_{+}(x,a_1) \equiv \mathcal{H}_2 \\ \notag &
\mathcal{D}_2=\mathcal{A}(x,a_0): \mathcal{H}_{-}(x,a_0)\equiv \mathcal{H}_2\rightarrow \mathcal{H}_{+}(x,a_0)\equiv \mathcal{H}_1 \\ \notag &
\mathcal{D}_3: \mathcal{H}_{-}(x,a_1) \equiv \mathcal{H}_1\rightarrow \mathcal{H}_{-}(x,a_1) \equiv \mathcal{H}_1
\end{align}
where in the third relationship, we see how $\mathcal{D}_3$ acts as a automorphism of the space $\mathcal{H}_{-}(x,a_1) \equiv \mathcal{H}_1$. To put it more simply, the Hilbert spaces $\mathcal{H}_1,\mathcal{H}_2,\mathcal{H}_3$ are identified with the Hilbert spaces of the shape invariant quantum systems in the following way:
\begin{align}\label{fgf}
&\mathcal{H}_1\equiv \mathcal{H}_{-}(x,a_1) \\ \notag &
\mathcal{H}_2\equiv \mathcal{H}_{-}(x,a_0) \\ \notag &
\mathcal{H}_3\equiv \mathcal{H}_{-}(x,a_1)\equiv \mathcal{H}_1
\end{align}
We easily obtain the operator $\mathcal{Q}$ and the Hamiltonian $H$ in terms of the operators $\mathcal{A}(x,a_1)$ and $\mathcal{A}(x,a_0)$,
\begin{equation}\label{grad1lphaop}
\mathcal{Q}= \left ( \begin{array}{ccc}
  0 & 0 & \mathcal{A}^{\dag}(x,a_1)\mathcal{A}^{\dag}(x,a_0) \\
  \mathcal{A}(x,a_1) & 0 & 0  \\
0 & \mathcal{A}(x,a_0)  & 0 \\
\end{array} \right)
\end{equation}
and
\begin{equation}\label{kdff}
H=\left ( \begin{array}{ccc}
  H_1 & 0 & 0  \\
  0 & H_2 & 0  \\
0 & 0 & H_3 \\
\end{array} \right)
\end{equation}
with the sub-Hamiltonians $H_1,H_2,H_3$ being equal to:
\begin{align}\label{subhamdifoff}
& H_1=\mathcal{A}(x,a_0)\mathcal{A}(x,a_1)\mathcal{A}^{\dag}(x,a_1)\mathcal{A}^{\dag}(x,a_0)\\ \notag &
H_2=\mathcal{A}(x,a_1)\mathcal{A}^{\dag}(x,a_1)\mathcal{A}^{\dag}(x,a_0)\mathcal{A}(x,a_0) \\ \notag &
H_3=\mathcal{A}^{\dag}(x,a_1)\mathcal{A}^{\dag}(x,a_0) \mathcal{A}(x,a_0)\mathcal{A}(x,a_1)
\end{align}
As we already mentioned, the $Z_3$-graded symmetric topological structure is actually guaranteed when the constraint (\ref{br}) is satisfied. When we substitute the operators $\mathcal{D}_i$ in terms of the operators $\mathcal{A}(x,a_i)$, the constraint (\ref{br}) becomes:
\begin{equation}\label{brtrans}
[\mathcal{A}(x,a_1)\mathcal{A}^{\dag}(x,a_1),\mathcal{A}^{\dag}(x,a_0)\mathcal{A}(x,a_0)]=0
\end{equation}
But, owing to the shape invariance relation (\ref{sic}), the condition (\ref{brtrans}) is automatically satisfied, since the function $f(a_0)$ is a constant number and is independent of $x$. This independence of the function $f(a_0)$ of $x$, is of particular importance in order to prove that the shape invariance condition implies the relation (\ref{brtrans}). Therefore, owing to relation (\ref{brtrans}), the quantum system defined by the identifications (\ref{atlanteansword}) and (\ref{fgf}) satisfies the constraints (\ref{cond1}) and (\ref{cond2}) and therefore constitutes a $Z_3$-graded symmetric quantum system, with $\mathcal{Q}$ and $H$, given by relations (\ref{grad1lphaop}) and (\ref{kdff}). The constraints (\ref{go}), (\ref{qcom}) and the algebra (\ref{topolsymmetrop}) are then trivially satisfied, as long as $\mathcal{K}=H$ holds true.

\noindent It worths if we summarize in short our results at this point. Every supersymmetric quantum mechanical system along with its shape invariant system can constitute a $Z_3$-graded quantum system, with the Hilbert space maps being those of relation (\ref{atlanteansword}). The only assumption we made is that the Hilbert space $\mathcal{H}_1$ is actually identical to the Hilbert space $\mathcal{H}_3$, so that in the end the operator $\mathcal{D}_3$ is just an automorphism. Owing to the shape invariance conditions (\ref{sic}), the constraint (\ref{br}) is trivially satisfied and the total quantum system consisting of the operators (\ref{atlanteansword}), (\ref{grad1lphaop}), (\ref{kdff}), satisfy the constraints (\ref{go}), (\ref{qcom}) and the algebra (\ref{topolsymmetrop}). Hence, the shape invariant supersymmetric quantum mechanical system can constitute a $Z_3$-graded symmetric topological quantum system. The arguments of the topological invariants that were used in reference \cite{ali1}, can also be used in the present case too, but these reduce to the simple Witten indices of the shape invariant systems. In addition we can also define the corresponding Betti number and spin complexes but we refrain going into details, since these results easily follow from the conclusions of reference \cite{ali1}. We were mostly interested to demonstrate the existence of the inherent $Z_3$ structure to the shape invariant sub-systems.

\subsection{A Sequence of Shape Invariant Supersymmetric Systems and the $Z_3$-graded Symmetry}

Simple one dimensional $N=2$ supersymmetry relates only a pair of Hamiltonians and regarding the non-zero modes, these Hamiltonians are isospectral \cite{reviewsusyqm}. In the case of shape invariant supersymmetric systems, shape invariance leads to sequence of Hamiltonians which are pairwise supersymmetric \cite{reviewsusyqm,shapeinvarianceapplications,bazeia}. In that case, the sequence of pairwise Hamiltonians looks like \cite{bazeia}:
\begin{align}\label{sequence}
&H^{(0)}=H_{-}(a_0)\\ \notag &
H^{(1)}=H_{+}=H_{-}(a_1)+R(a_1)\\ \notag &
H^{(2)}=H_{-}(a_2)+R(a_1)+R(a_2) \\ \notag &
H^{(k)}=H_{-}(a_k)+\sum_{j=1}^{k}R(a_j)
\end{align}
It is more convenient for our purposes to write relation (\ref{sequence}) in terms of the corresponding operators $\mathcal{A}(x,a_i)$. So writing the Hamiltonians of relation (\ref{sequence}) in terms of the $\mathcal{A}$ operators, we get:
\begin{align}\label{alphaops}
&\mathcal{A}(x,a_0)\mathcal{A}^{\dag}(x,a_0)=\mathcal{A}^{\dag}(x,a_1)\mathcal{A}(x,a_1)+R(a_1)\\ \notag &
\mathcal{A}(x,a_1)\mathcal{A}^{\dag}(x,a_1)=\mathcal{A}^{\dag}(x,a_2)\mathcal{A}(x,a_2)+R(a_1)+R(a_2) \\ \notag &
\mathcal{A}(x,a_{k-1})\mathcal{A}^{\dag}(x,a_{k-1})=\mathcal{A}^{\dag}(x,a_k)\mathcal{A}(x,a_k)+\sum_{j=1}^{k}R(a_j)
\end{align}
Following the line of research of the previous section, it is obvious that each shape invariant subsystem characterized with the variables $(a_{k-1},a_k)$, with $k=0,1,...k$, can solely constitute a $Z_3$-graded quantum mechanical system. Indeed, the $Z_3$-graded symmetric structure with parameters $(a_{k-1},a_k)$, has the following Hamiltonian and $\mathcal{Q}$ operator:
\begin{equation}\label{grad1lphaogfgfgp}
\mathcal{Q}= \left ( \begin{array}{ccc}
  0 & 0 & \mathcal{A}^{\dag}(x,a_k)\mathcal{A}^{\dag}(x,a_{k-1}) \\
  \mathcal{A}(x,a_k) & 0 & 0  \\
0 & \mathcal{A}(x,a_{k-1})  & 0 \\
\end{array} \right)
\end{equation}
and
\begin{equation}\label{kdffddsfdff}
H^{k}=\left ( \begin{array}{ccc}
  H_1^{k} & 0 & 0  \\
  0 & H_2^{k} & 0  \\
0 & 0 & H_3^{k} \\
\end{array} \right)
\end{equation}
with the sub-Hamiltonians $H_1,H_2,H_3$ being equal to:
\begin{align}\label{subhamdiffgfgfoff}
& H_1^{k}=\mathcal{A}(x,a_{k-1})\mathcal{A}(x,a_k)\mathcal{A}^{\dag}(x,a_k)\mathcal{A}^{\dag}(x,a_{k-1})\\ \notag &
H_2^{k}=\mathcal{A}(x,a_k)\mathcal{A}^{\dag}(x,a_k)\mathcal{A}^{\dag}(x,a_{k-1})\mathcal{A}(x,a_{k-1}) \\ \notag &
H_3^{k}=\mathcal{A}^{\dag}(x,a_k)\mathcal{A}^{\dag}(x,a_{k-1}) \mathcal{A}(x,a_{k-1})\mathcal{A}(x,a_k)
\end{align}
For $k=0,1,...,k$, each of the operators $\mathcal{A}(x,a_i)$, participates to two different $Z_3$ graded symmetric quantum mechanical system, except for $\mathcal{A}(x,a_0)$ and $\mathcal{A}(x,a_k)$, which participate in only one. Hence, for a sequence of $k+1$ operators,
\begin{equation}\label{dhdhsucks}
\mathcal{A}(x,a_{0}),\mathcal{A}(x,a_{1}),\mathcal{A}(x,a_{2}),...,\mathcal{A}(x,a_{k-1}),\mathcal{A}(x,a_{k}),
\end{equation}
a number of $k$ different $Z_3$-graded symmetric systems can be constructed. Finally, before we proceed to the next section, let us comment that there is no obvious connection between the $k$ different $Z_3$-graded algebras. This requires somewhat more detailed study, which stretches beyond the purposes of this article.

\section{Connection of the $Z_3$-graded Topological Symmetry with non-linear realizations of Lie Algebras}

In this section, owing to the fact that the shape invariant SUSY QM systems can be associated to a $Z_3$ quantum symmetry, we shall relate the $Z_3$ system with an $SO(2,1)$ Lie algebra. As is already established in the literature \cite{bala1}, every shape invariant supersymmetric quantum mechanical system has an underlying $SO(2,1)$ Lie algebraic structure. Having in mind the shape invariance equation (\ref{sic}), let us recall how this algebra is constructed. Introducing an auxiliary variable $\phi$, we define the operators $J_+$ and $J_{-}$ as follows \cite{bala1}:
\begin{equation}\label{operlie}
\mathcal{J}_{+}=Q(\phi)\mathcal{A}^{\dag}(x,\chi (i\partial_{\phi})),{\,}{\,}{\,}\mathcal{J}_{-}=\mathcal{A}(x,\chi (i\partial_{\phi}))Q*(\phi )
\end{equation}
with $Q(\phi )$ a function to be determined and $\chi (i\partial_{\phi})$ an arbitrary function. Recall the analytic form of the operators $\mathcal{A}$ and $\mathcal{A}^{\dag}$, which we now give for convenience:
\begin{equation}\label{opeaanlytic}
\mathcal{A}(x,a)=\frac{\mathrm{d}}{\mathrm{d}x}+W(x),{\,}{\,}{\,}\mathcal{A}^{\dag}(x,a)=-\frac{\mathrm{d}}{\mathrm{d}x}+W(x)
\end{equation}
The operators $\mathcal{A}^{\dag}(x,\chi (i\partial_{\phi}))$ and $\mathcal{A}(x,\chi (i\partial_{\phi}))$ appearing in (\ref{operlie}), are obtained from the ones in (\ref{opeaanlytic}), by the substitution $a\rightarrow \chi (i\partial_{\phi})$. Finding the commutator of the operators $\mathcal{J}_{+}$ and $\mathcal{J}_{-}$, with $Q(\phi )=e^{ip\phi}$ (where $p$ is an arbitrary real constant) and bearing in mind the shape invariance condition becomes:
\begin{equation}\label{siclie}
\mathcal{A}(x,\chi (i\partial_{\phi}))\mathcal{A}^{\dag}(x,\chi (i\partial_{\phi}))-\mathcal{A}^{\dag}(x,\chi (i\partial_{\phi}+p))\mathcal{A}(x,\chi (i\partial_{\phi}+p))=f(\chi (i\partial_{\phi}))
\end{equation}
we obtain the deformed $SO(2,1)$ Lie algebra underlying the shape invariant systems, that is:
\begin{align}\label{liedeform}
& [\mathcal{J}_3,\mathcal{J}_{\pm}]=\pm\mathcal{J}_{\pm}, \\ \notag &
[\mathcal{J}_+,\mathcal{J}_{-}]=\xi (\mathcal{J}_3)
\end{align}
In the above, $\mathcal{J}_3$ stands for $\mathcal{J}_3=-i\partial_{\phi}/p$ and $\xi (\mathcal{J}_3)=-f(\chi (i\partial_{\phi}))$. Due to the existing structure, the $Z_3$-graded symmetry we studied in the previous section can be written in terms of the operators of the Lie algebra. However, the cases in which this can be done are somewhat more restricted than in the previous section case. Actually the $Z_3$-graded symmetric structure always exists when,
\begin{equation}\label{hfgffd}
[W(x,\chi (i\partial_{\phi}),i\chi (i\partial_{\phi})]=0
\end{equation}
Now it is time to specify what forms the function $\chi (i\partial_{\phi})$ can take. These forms of the function are determined from the particular shape invariance of the system. We shall mainly be interested in the cases of translational and multiplicative shape invariance, for which the function $\chi$ is $\chi(z)=z$ and $\chi(z)=e^z$ respectively. For these cases, when the potential function $W(x,\chi (i\partial_{\phi})$ is a linear operator with respect to the operator $\chi (i\partial_{\phi})$, relation (\ref{hfgffd}) holds true. This is because the following commutators are zero:
\begin{equation}\label{commzero}
[\partial_{\phi},\chi]=0,{\,}{\,}{\,}[\frac{\mathrm{d}}{\mathrm{d}x},\partial_{\phi}]=0
\end{equation}
Owing to relation (\ref{commzero}) and when relation (\ref{hfgffd}) holds true, the $Z_3$-graded symmetry of the previous section can be written in terms of the generators of the $SO(2,1)$ Lie algebra. Indeed, the operator $\mathcal{Q}$ and the Hamiltonian $H$ of the $Z_3$ graded system can be written as follows:
\begin{equation}\label{liegrad1lphaop}
\mathcal{Q}= \left ( \begin{array}{ccc}
  0 & 0 & \mathcal{J}_{+}e^{-ip\phi}\mathcal{J}_{+} \\
  \mathcal{J}_{+} & 0 & 0  \\
0 & e^{ip\phi}\mathcal{J}_{-} & 0 \\
\end{array} \right)
\end{equation}
and
\begin{equation}\label{liekdff}
H=\left ( \begin{array}{ccc}
  H_1 & 0 & 0  \\
  0 & H_2 & 0  \\
0 & 0 & H_3 \\
\end{array} \right)
\end{equation}
with the sub-Hamiltonians $H_1,H_2,H_3$ being equal to:
\begin{align}\label{liesubhamdifoff}
& H_1=e^{ip\phi}\mathcal{J}_{-}\mathcal{J}_{-}\mathcal{J}_{+}e^{-ip\phi}\mathcal{J}_{+}\\ \notag &
H_2=\mathcal{J}_{-}\mathcal{J}_{+}e^{-ip\phi}\mathcal{J}_{+}e^{ip\phi}\mathcal{J}_{-} \\ \notag &
H_3=\mathcal{J}_{+}e^{-ip\phi}\mathcal{J}_{+} e^{ip\phi}\mathcal{J}_{-}\mathcal{J}_{-}
\end{align}
It is obvious that the constraint (\ref{br}) is automatically satisfied when relation (\ref{hfgffd}), owing to relation (\ref{siclie}) holding true. In conclusion, the inherent $SO(2,1)$ Lie algebraic structure to any shape invariant supersymmetric quantum mechanical system, can be used to produce a $Z_3$-graded symmetric quantum system, with Hamiltonian (\ref{liekdff}) and $\mathcal{Q}$-operator given by (\ref{liegrad1lphaop}). We can go further and express the $Z_3$-graded elements in terms of creation and annihilation operators related to shape invariant systems, but since the results are similar to those presented in this section we refrain from going into details. The interested reader can convince himself by using the results of reference \cite{bala1} and apply the techniques along the lines of argument we used in this section.

\subsection{A Brief Comment on Shift Operators, Shape Invariance and the $Z_3$ Symmetry}

It is believed in the literature \cite{spector}, that extended SUSY QM with central charge (SQMCC hereafter) is a stepping stone to shape invariance. Bearing this in mind, and the results we established in the previous sections, that is, every shape invariant supersymmetric quantum system can be related to an extended $Z_3$-graded symmetric system, in this section the focus is on the possible connection of the centrally extended SUSY QM systems with the $Z_3$ symmetry. Particularly, we shall be interested in the qualitative features of the aforementioned issues and we defer the quantitative issues for a future article, more focused on the subject.

As is shown in reference \cite{spector}, when the algebra of SUSY QM algebra with central charge is enhanced by an additional symmetry, shape invariance emerges in a natural way. Let us briefly present the features of this extra symmetry and outline the lines along with one could find a possible connections between the two symmetries. Following \cite{spector}, when we take SQMCC and add an additional symmetry condition, namely that the so-called shift operator $\mathcal{S}$, is conserved, that is it commutes with the Hamiltonian, $[H,\mathcal{S}]=0$. The operator $\mathcal{S}$ has the representation,
\begin{equation}\label{shapeinvliekdff}
\mathcal{S}=\left ( \begin{array}{cccc}
  0 & 0 & 0 & 0 \\
  A_1 & 0 & 0 & 0 \\
0 & C & 0 & 0\\
0 & 0 & A_3 & 0\\
\end{array} \right)
\end{equation}  
with $A_1$ and $A_3$ operators related to the SQMCC algebra and similar to those of relation (\ref{opeaanlytic}). When does an operator $\mathcal{S}$ exists, translates to finding the suitable operator $C$. As pointed out in \cite{spector}, a conserved $\mathcal{S}$ can be defined when the following condition holds true,
\begin{equation}\label{hdggdfugf}
A_1(g_1)A_1^{\dag}(g_1)=A_1^{\dag}(g_2)A_1(g_2)+f(g_1)
\end{equation}
which is nothing else but the shape invariance condition for the SUSY QM subsystem $A_1$. In addition, the operator $A_3$ is obtained by a two step iteration of the shape invariance transformation on $A_3$. Moreover, the operator $C$ is obtained by one more iteration on the shape invariance condition of $A_1$. Hence, shape invariance is a result of the existence of this operator. In view of the previous sections results, the question is how this operator can be related to the $Z_3$-graded symmetric algebra and in particular, due to the condition $[\mathcal{Q},H]=0$, what is the algebraic relation (if any) between $\mathcal{Q}$ and $\mathcal{S}$. Having in mind that the shape invariance condition is usually imposed as an ad hoc relation, without having a structural-algebraic reason, the possible connection between the $\mathcal{S}$-symmetry and the $Z_3$-symmetry, might shed some light on the structural-algebraic origin of shape invariance. Although interesting, such a task is beyond the scope of this article, so we defer this work to a more focused article on this problem.

\section*{Concluding Remarks}

In this article we studied shape invariant SUSY QM systems and demonstrated that each shape invariant system can constitute a $Z_3$-graded symmetric quantum mechanical system. Particularly, the algebra of $Z_3$-graded quantum symmetry leads to some constraints which need to be satisfied by all the elements \cite{ali1}. We chose the solution of \cite{ali1}, which leads to another set of constraints. As we explicitly showed, each shape invariant SUSY QM system automatically satisfies these constraints. This result can be generalized to take into account a system of sequential shape invariant systems, in which case we found that each system can constitute a different $Z_3$-graded symmetry. Furthermore, since each shape invariant system has an inherent $SO(2,1)$ Lie algebra, we related the generators of the Lie algebra to the $Z_3$-graded symmetry. Finally, we presented two systems that are related in an algebraic way to shape invariance, namely a supersymmetric quantum system with central charge and with an additional symmetry and the $Z_3$-graded symmetric systems. Owing to the fact that the shape invariance condition is imposed as an extra condition, without referring to any algebraic reasoning, this similarity between the $\mathcal{S}$-symmetry and the $Z_3$-symmetry, might enlighten the problem of finding a structural-algebraic origin of shape invariance. This task certainly deserves to be studied in a future work. Moreover, since each $Z_3$-graded symmetric system has a cohomological structure, our results might be useful from a mathematical point of view, since each shape invariant system is related to a cohomological structure and in addition, there exists a generalized Witten index which is directly related to the Witten index of the shape invariant system. Certainly, such issues deserve some attention and we hope to address such issues in a future work.

\end{document}